\documentclass[%
twocolumn,
superscriptaddress,
 amsmath,amssymb,
 aps,
]{revtex4}

\usepackage{graphicx}
\usepackage{dcolumn}
\usepackage{bm}

\usepackage{textgreek} 
\usepackage{xcolor}
\usepackage{units}

\begin{document}

\preprint{APS/123-QED}

\title{Elastic stresses reverse Ostwald ripening}

\author{Kathryn A. Rosowski}
 \affiliation{Department of Materials, ETH Z\"{u}rich, 8093 Z\"{u}rich, Switzerland.}
 
\author{Estefania Vidal-Henriquez}
\affiliation{Max Planck Institute for Dynamics and Self-Organization, 37077, G\"{o}ttingen, Germany}

\author{David Zwicker}
\affiliation{Max Planck Institute for Dynamics and Self-Organization, 37077, G\"{o}ttingen, Germany}

\author{Robert W. Style}
 \affiliation{Department of Materials, ETH Z\"{u}rich, 8093 Z\"{u}rich, Switzerland.}
 
\author{Eric R. Dufresne}
 \email{eric.dufresne@mat.ethz.ch}
 \affiliation{Department of Materials, ETH Z\"{u}rich, 8093 Z\"{u}rich, Switzerland.}

\date{\today}

\begin{abstract}
When liquid droplets nucleate and grow in a polymer network, compressive stresses can significantly increase their internal pressure, reaching values that far exceed the Laplace pressure.
When droplets have grown in a polymer network with a stiffness gradient, droplets in relatively stiff regions of the network tend to dissolve, favoring growth of droplets in softer  regions.
Here, we show that this elastic ripening can be strong enough to reverse the direction of Ostwald ripening:  large droplets can shrink to feed the growth of smaller ones.
To numerically model these experiments, we generalize the theory of elastic ripening to account for gradients in solubility alongside gradients in mechanical stiffness.
\end{abstract}

\maketitle


\section{Introduction}

\begin{figure}
\centering
  \includegraphics[width=7cm]{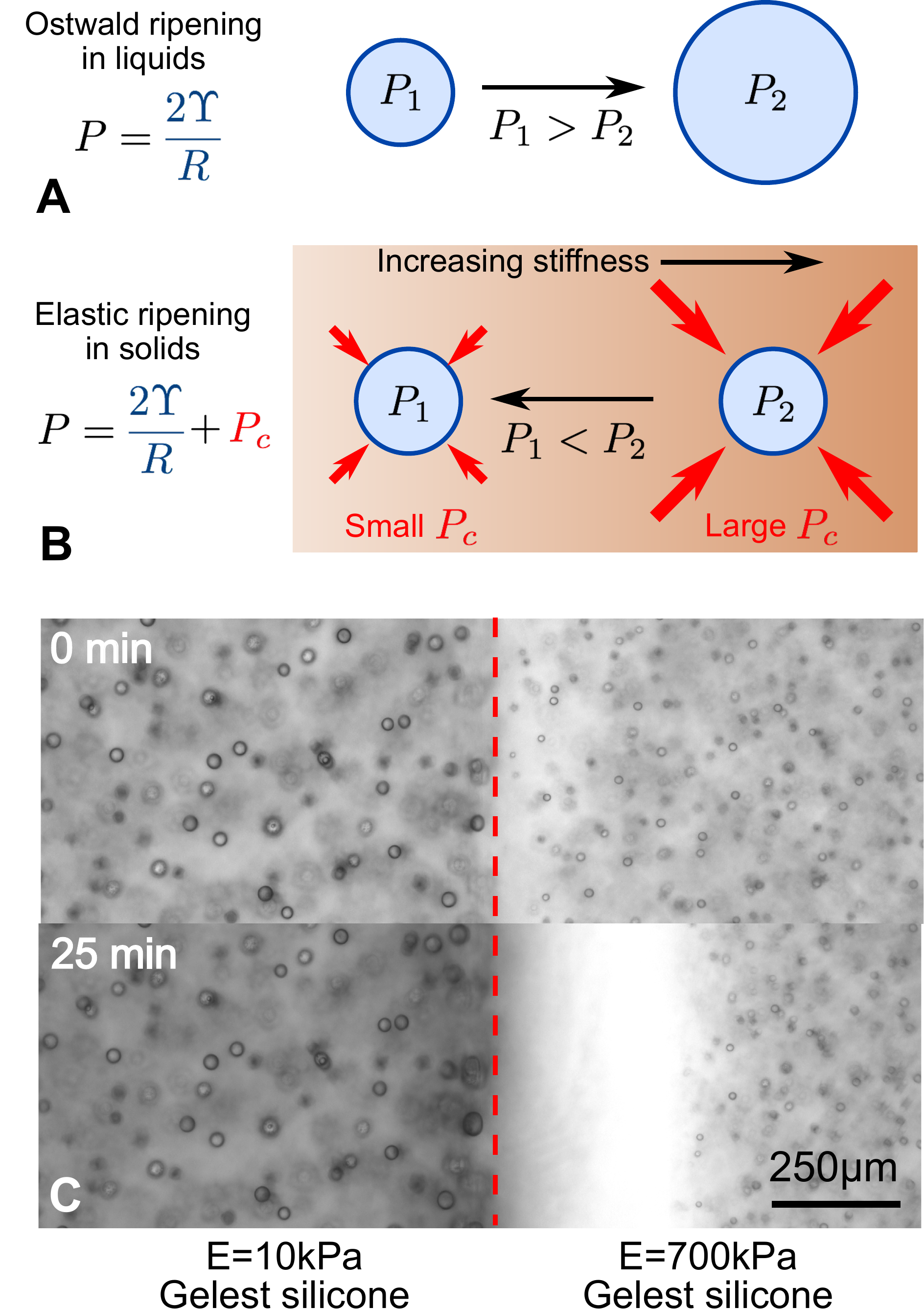}
  \caption{\emph{Ostwald ripening in liquids and elastic ripening in polymer networks}. A,B) Schematic diagrams summarizing the key differences between Ostwald ripening and elastic ripening \cite{roso2020}. Here, $P_c\sim E$ is the cavitation pressure. C) In substrates with a gradient in stiffness (here a step change from $E=10-700$kPa), but with otherwise identical properties, droplets ripen from the stiffer side to the softer side (data from \cite{roso2020}).}
  \label{fig:schem}
\end{figure}

In a conventional emulsion, the long term-stability of droplets is typically limited by their interfacial energy.
Over time, the size distribution of droplets coarsens, with smaller drops disappearing in favor of larger ones \cite{dege04}.
The fastest route to coarsening is the direct coalescence of droplets.
However, when coalescence is suppressed -- typically by surfactants -- Ostwald ripening takes over \cite{cates2017complex}.
In this process, summarized in Figure \ref{fig:schem}a, small droplets tend to shrink by dissolution and large ones tend to grow by condensation.
This is driven by differences in the droplets' Laplace pressure, $P=2\Upsilon/R$, where $\Upsilon$ is the surface tension and $R$ is the droplet radius.

This picture changes significantly when droplets form by nucleation and growth in a polymer network.
These droplets tend to be monodisperse, with smaller droplets appearing in stiffer networks \cite{styl2018}.  
When droplets exclude the polymer network, they push the network outward as they grow.
In response, the network squeezes the droplets.
This both suppresses droplet condensation \cite{roso2020}, and increases the droplet's internal pressure by an amount comparable to the network's Young modulus, $E$ \cite{hutchens2016elastic,raayai2019intimate,Kim2020}.
This increase in pressure can potentially far exceed the Laplace pressure.
Thus, when the polymer network has heterogeneous mechanical properties, the elastic contribution to droplet pressure is heterogeneous and can drive transport of  material from droplets in stiff regions to droplets in soft regions, summarized in Figure \ref{fig:schem}b \cite{roso2020}. 
Like Ostwald ripening, this `elastic ripening' is mediated by the transport of material between droplets in the dilute phase.
Related phenomena have been observed in the nucleus of living cells \cite{shin2018}.

Previous demonstrations of elastic ripening,  like that in Figure \ref{fig:schem}c, have superficially resembled Ostwald ripening, because large droplets (in soft regions of the network), grew at the expense of small droplets (in stiff regions of the network).
Thus, the establishment of elastic stresses as the driving force for ripening relied on comparisons across experiments with varying elastic moduli, and comparisons to numerical models.

Here, we demonstrate that elastic ripening phenomena can qualitatively differ from classic Ostwald ripening.
In particular, we demonstrate the growth of small droplets, fed by the dissolution of large ones.
Experimentally, this is achieved by coupling two different families of silicone gels, which have different thermodynamic and transport properties at the same elastic modulus.
These new results demand a generalization of the model for elastic ripening. 
In particular,  gradients in solubility must be accounted for.
With this generalization,  numerical simulations capture essential features of experimental observations, using experimentally measured properties.

\section{Experimental Results}

We study elastic ripening in a system of phase-separated fluorinated oil (Fluorinert FC-770) droplets in silicone gels.  
The gels are saturated in a bath of fluorinated oil at 40$^\circ$C over a few days,  then cooled passively to room temperature (22-23$^\circ$C) over several minutes.
As the samples cool, the solubility of the fluorinated oil decreases, and droplets grow in the gels \cite{styl2018}.
The silicone network is excluded from the droplets, so they grow by pushing open holes in the gel \cite{Kim2020}.

To create stiffness gradients, we make two different silicones side by side.
We first cure a 3-5mm layer of the stiffer silicone in a polystyrene petri dish (Greiner).
Half of this is cut out with a razor blade, pulled off the dish, and placed in half of a new, glass-bottomed dish (MatTek).
The softer silicone is then poured into the other side of the dish and allowed to cure \cite{roso2020}.
When we drive phase separation, droplets on each side of the sample grow to a uniform distribution, with a sharp transition between them (Figure \ref{fig:schem}c).
After relatively fast droplet formation, we observe slow evolution of the droplets near the interface.

In previous experiments \cite{roso2020}, we observed that smaller droplets, on the stiffer side, dissolved and fed the growth of larger droplets, on the softer side (e.g. Figure \ref{fig:schem}c).
While resembling familiar Ostwald ripening, due to surface tension, we concluded that ripening was driven by gradients in the stiffness of the polymer network.  
This conclusion was based on an observed increase of the coarsening rate with the stiffness difference and comparison with numerical models. 
Here, we challenge this interpretation by considering ripening across a broader range of silicone samples.

We use two different families of silicone gels, as described in the Materials and Methods.
`Gelest' silicones are fabricated by mixing together functionalized silicone polymers, crosslinker, and a platinum catalyst following \cite{roso2020}.
`Sylgard' silicones are fabricated with a popular commercial kit, Sylgard 184.  
In addition to silicone polymer, this contains a significant quantity of silica nanoparticle filler  \cite{lee2004}.
With both types of silicone, we can achieve a range of Young's moduli, $E$, from tens to hundreds of kPa.

\begin{figure}
\centering
  \includegraphics[width=7cm]{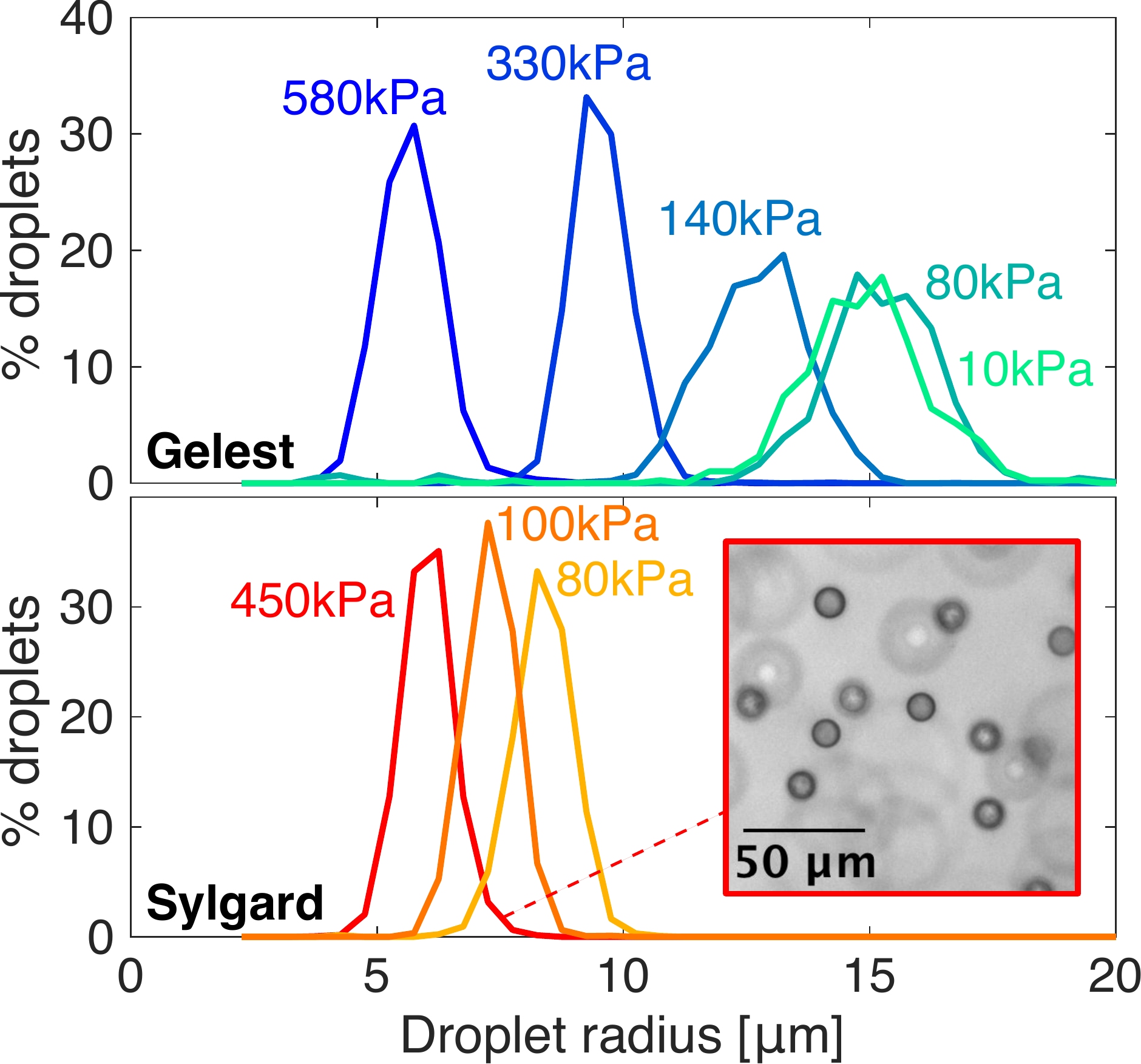}
  \caption{\emph{Droplets in Sylgard and Gelest silicones have different sizes}. The size distributions of droplets formed by phase separation in different stiffness silicones. For a given stiffness, droplets in Gelest silicones are typically larger than droplets in Sylgard silicones. The inset shows a typical image of droplets formed in Sylgard silicone with $E=450$kPa. }
  \label{fig:sizes} 
\end{figure}

Despite some similarities, these two families of silicones show quantitative differences in their phase-separation behavior.
In particular, condensed droplets of fluorinated oil have different sizes at the same network stiffness, as shown in Figure \ref{fig:sizes}.
As the stiffness of Gelest networks increases from 10 to 580 kPa, the mean droplet radius reduces from $14.9\pm 1.4\mu\mathrm{m}$ to $5.7\pm 0.7\mu\mathrm{m}$.  
As the stiffness of the Sylgard networks increases from 80 to 450 kPa, the mean droplet radius reduces slightly from  $8.3\pm 0.6\mu\mathrm{m}$ to $6.0\pm 0.5\mu\mathrm{m}$.

In samples made from only one of these silicone families,  elastic ripening and Ostwald ripening proceed in the same direction. 
This is because the droplet size generically decreases with the stiffness.
With two different families of silicone, we can remove this limitation,
and explore combinations of materials where elastic and Ostwald ripening alternatively reinforce or oppose each other.

\begin{figure}
 \centering
 \includegraphics[width=8.5cm]{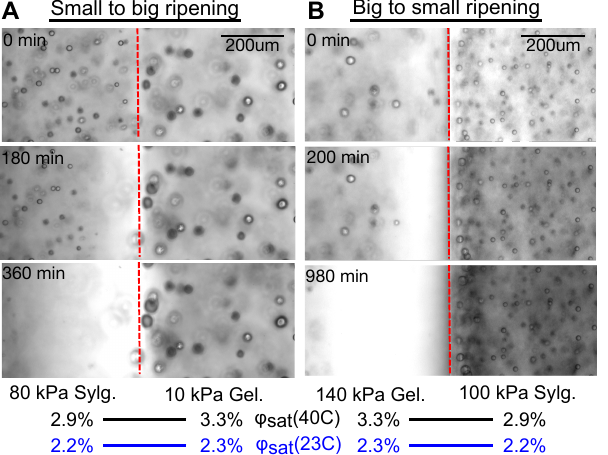}
 \caption{\emph{Elastic ripening can go against Ostwald ripening.} A) A gradient consisting of stiff (40 kPa) Sylgard and soft (10kPa) Gelest displays movement from stiff to soft, corresponding to movement of material from small to large droplets.  B) A gradient consisting of stiff (140 kPa) Gelest and soft (40kPa) Sylgard displays movement from stiff to soft, corresponding to movement of material from large to small droplets. The image contrasts have been adjusted for clarity. The red dashed line indicates the interface between the two gels.}
 \label{fig:ripening}
\end{figure}

An example where Ostwald and elastic ripening reinforce each other is shown in Figure \ref{fig:ripening}a and Supplemental Movie 1.
Here, stiffer Sylgard silicone ($E=80$kPa, mean droplet size 8.4$\mu$m) is in contact with softer Gelest silicone ($E=10$kPa, mean droplet size 14.9$\mu$m).
Consistent with previous results, smaller fluorinated oil droplets on the stiff side shrink while feeding the growth of larger droplets on the soft side.
The direction of ripening is consistent with both Ostwald and elastic ripening.

An example where elastic forces oppose Ostwald ripening is shown in Figure \ref{fig:ripening}b and Supplemental Movie 2.
Here, stiff Gelest silicone ($E=140$kPa, mean droplet size 12.7$\mu$m) is in contact with soft Sylgard silicone ($E=100$kPa, mean droplet size 7.3$\mu$m).
In this case, the larger droplets near the interface on the stiff side shrink while small droplets on the soft side grow.
This is the opposite of the trend  expected from Ostwald ripening, and provides a convincing visual case for elastic ripening.

\begin{figure}
\centering
  \includegraphics[width=7cm]{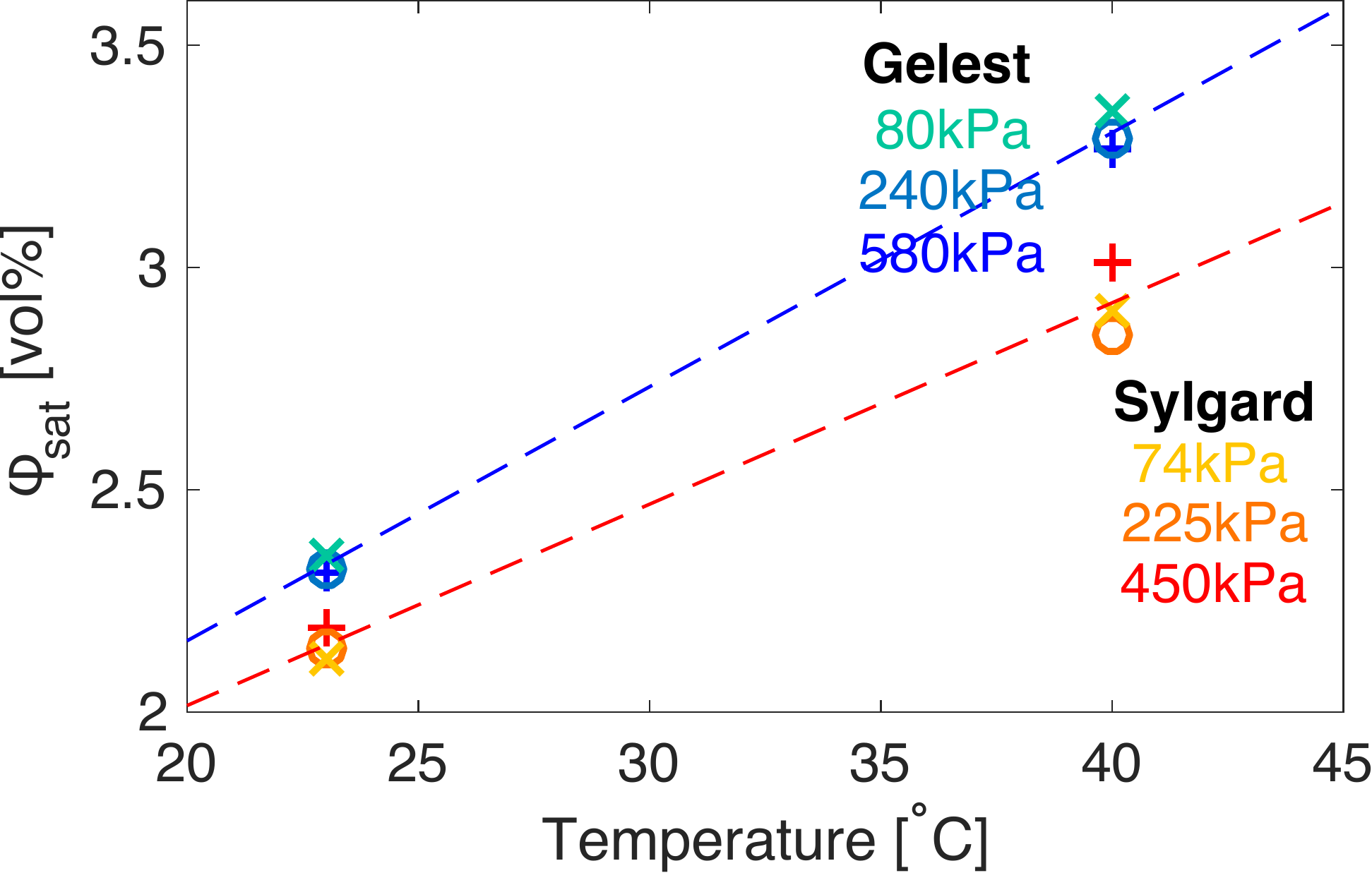}
  \caption{\emph{Sylgard and Gelest silicones have different saturations}.  A plot of $\phi_{sat}$ as a function of temperature shows how fluorinated oil is more soluble in Gelest than in Sylgard silicones. The solubility is effectively independent of stiffness for the two different types of silicone. }
  \label{fig:phi_sat} 
\end{figure}

An interesting complication in the interpretation of the data in Figure \ref{fig:ripening} arises due to the differences in the solubility of fluorinated oil on the two sides.  
In both silicones, the saturation concentration, $\phi_{sat}$, increases with the temperature, but does not change significantly with the elastic modulus (Figure \ref{fig:phi_sat}).
However, fluorinated oil has significantly higher solubility in the Gelest than Sylgard silicones, Figure \ref{fig:phi_sat}.
Additionally, the solubility of fluorinated oil is more temperature sensitive in Gelest than in Sylgard.
In practice, this means that when the temperature drops, the Gelest samples become more strongly supersaturated than the Sylgard ones.
As we will see below, these factors play important roles in  understanding the observed ripening behaviour.

\section{Theory and Simulation}

In the classic description of Ostwald ripening \cite{voorhees1985} and initial descriptions of elastic ripening \cite{vida2020}, the concentration, $\phi$, of solute near the surface of a droplet is pinned to its equilibrium value, which increases with the internal pressure of the droplet. 
Then, simple diffusion drives a flux $\vec J$ of solute down concentration gradients according to Fick's law, $\vec J=-D \vec \nabla \phi$, where $D$ is the diffusivity of the oil in the dilute phase.

In our experiments, solute transport occurs in gels with a heterogeneous saturation concentration.
In such cases, one needs to take a more general approach, where fluxes are driven by gradients in chemical potential, $\mu$:
\begin{equation}
\vec J=-\frac{\phi D}{k_BT} \vec \nabla \mu.
\label{eqn:J}
\end{equation}
Here, $k_B$ is Boltzmann's constant, and $T$ is temperature.
Using dilute solution theory, the chemical potential of a solute can be approximated as  \cite{styl2018,vida2020}:
\begin{equation}
    \mu=k_BT\log(\phi/\phi_{sat}).
    \label{eqn:mu}
\end{equation}
Note that there are a range of more complex expressions for $\mu$ that capture more of the physics of the polymer network -- for example
Flory-Rehner theory \cite{flory1943statistical} -- however this simple model captures the key physics.
When $\phi_{sat}$ is homogeneous, Eqs. \ref{eqn:J} and \ref{eqn:mu} reduce to Fick's law.
Howevever, when gradients in $\phi_{sat}$ are signficant, they can dominate over gradients in concentration.

In the droplet phase, the chemical potential is simply related to the pressure, $\mu\approx P/n_L$ \cite{liu2016}.  
Here, $n_L$ is the number density of molecules in the droplet phase.
Differences in chemical potential between droplets are thus equivalent to  pressure differences,  and
Eq. \ref{eqn:J} matches our  expectation that solute should be driven from high pressure to low pressure.

For droplets in an elastic network, the pressure within the droplet has the form,
\begin{equation}
    P = \frac{2\Upsilon}{R} + P_c.
    \label{eqn:pressure}
\end{equation}
The first term is the contribution from surface tension, unchanged from classic Ostwald ripening.
The second term is the contribution from compressive stresses from the polymer network.
Due to the cavitation instability, $P_c=\alpha E$ is independent of droplet size and $\alpha$ is a constant close to one, describing the cavitation process \cite{gent91,zimb07,Kim2020,roso2020}.
Thus, the difference in pressure between droplets in the two domains is  $\Delta P=2\Upsilon\Delta R/R^2+\alpha \Delta E$, where $\Delta E$ and $\Delta R$ are the differences in Young's modulus and droplet radii between the two domains.
By comparing these terms, we find elastic-dominated ripening when
\begin{equation}
\left(\frac{\Upsilon}{\Delta E R}\right)\left(\frac{ \Delta R}{ R}\right)\ll1.
\end{equation}
For a silicone/fluorinated-oil interface, $\Upsilon=4\mathrm{mN/m}$, so for all droplets here $\Upsilon /(\Delta E R) \ll 1$.
Since $\Delta R/R$ is $O(1)$, elastic ripening dominates, (\emph{i.e.} $\Delta P \sim \Delta E$), and the ripening direction is solely determined by the stiffness difference, independent of the droplet size.

With this simple extension of ripening theory, numerical simulations capture essential features of the experiments, as shown in Figure \ref{fig:ripening}.
We randomly place droplets on either side of the sample, with distributions matching the measured experimental results.
Since the droplets equilibrate quickly with their surrounding, we use the chemical potential associated with the pressure $P$ given by Eq. \eqref{eqn:pressure} to calculate the concentration~$\phi$ using Eq. \eqref{eqn:mu}.
Finally, we model diffusive flux between droplets by combining equations (\ref{eqn:J},\ref{eqn:mu}) with a conservation law, as described in the Materials and Methods.
Values of $D$, $\phi_{sat}$ and $E$ are taken directly from the measured values.

\begin{figure}[tb]
 \centering
 \includegraphics[width=0.98\linewidth]{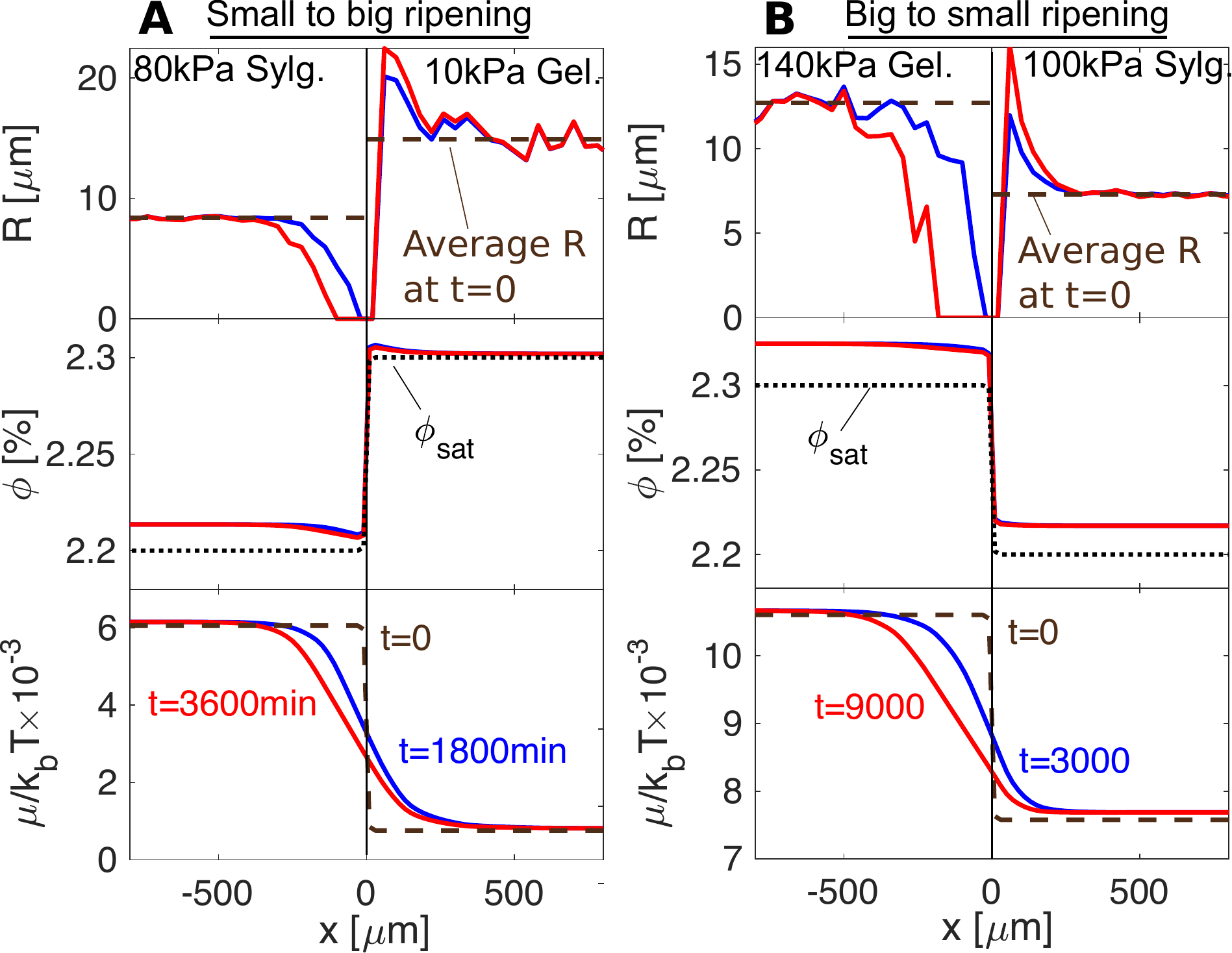}
 \caption{\emph{Numerical simulations of the experimental setups presented in Figure \ref{fig:ripening}}. Top: Droplet radius. Middle: Concentration in the dilute phase, $\phi$. Bottom: chemical potential, $\mu$. A) For 80kPa Sylgard next to 10kPa Gelest, elastic ripening moves against concentration gradients. B) For 140kPa next to 100kPa Sylgard, elastic ripening goes against classical Ostwald ripening.}
 \label{fgr:simulation}
\end{figure}

The results, shown in Figure \ref{fgr:simulation} and Supplemental Movies 3 and 4, are in  qualitative agreement with the experiments.
The left-hand column shows results of the simulation of the experiment of Figure \ref{fig:ripening}a.
As in the experiment, we see depletion of the droplets in the stiffer Sylgard, and growth of droplets in the softer Gelest adjacent to the interface.
Interestingly, we see that $\phi$ is always larger on the softer side than on the stiffer side, so diffusion progresses up concentration gradients.
This is due to the differences in $\phi_{sat}$ between the two sides.
When we account for this in the chemical potential, via equation (\ref{eqn:mu}), we verify that transport occurs down chemical potential gradients, as expected. 
The right-hand column of the figure simulates the experiment of Figure \ref{fig:ripening}b.
Again, we see the ripening from the stiff side to the soft side.
However, in this case, ripening occurs from larger droplets to smaller droplets -- the opposite direction to what is usually expected.

While the simulations capture the essential experimental trends, transport in experiments is approximately ten times faster than expected from simulations.
The origin of this discrepancy remains elusive.

\section{Conclusions}

We have demonstrated that elastic stresses in polymer networks can reverse the direction of droplet ripening.
To understand these results, we have generalized the theory of elastic ripening to account for simultaneous gradients of elasticity and solubility.
In these cases, the simple picture of diffusive transport down a concentration gradient must be set aside in favor of a more general approach, where transport occurs along gradients in the chemical potential.
Qualitatively, the direction of transport can be predicted simply by considering the internal pressure of the droplets.
Since the chemical potential of the pure liquid droplets increases with their pressure, oil is always transported from high pressure regions to low pressure regions.
For droplets grown in a polymer network, their pressure generally has contributions from both surface tension and compressive network stresses.  
For the cases here, the latter is much higher than the former.
Quantitative prediction of the rate of ripening further requires knowledge of the saturation concentrations and diffusion coefficients of dissolved oil in the networks.

These results help to lay the foundations for the analysis of phase separation in complex, heterogeneous environments.
Our experiments and theory are inspired by recent observations of phase separation in living cells (e.g. \cite{shin2018,quiroz2020liquid}).
In that context, specific macromolecules, including nucleic acids and proteins, segregate into functional domains within the cytoplasm and nucleoplasm \cite{alberti2016aberrant,bran09,hyma14,shin17}.
The working model for interpreting these phenomena is the phase separation of two component fluids.
While this captures some of the basic phenomenology, recent theories aimed to account for activity \cite{zwic14,webe19}, complex compositions \cite{jacobs2017}, and rheology \cite{tana00,styl2018,vida2020,hennessy2020phase,mukherjee2019gelation}.
Our work further establishes a general framework for evaluating the role of a passive network in crowded systems.
Note that we assume that our system remains near equilibrium and that the rheology of the continuous phase can be reduced to a single, static, inflation pressure.
Thus, these results should not be directly applied to complex living systems without caution.
We also anticipate that these results may be useful in designing other phase separation processes in materials near equilibrium, such as the production of porous membranes and scaffolds \cite{yang08} or the formation of segregated ice during the processing of frozen foods (e.g. \cite{van2006}), in cryopreservation \cite{karl96}, or in other processes where freezing causes material damage to porous materials (e.g. \cite{scho16}).

\section{Materials and Methods}
\subsection{Preparation of silicone gels}
To create `Gelest' silicones, we follow the recipe in \cite{styl15}, divinyl-terminated polydimethylsiloxane chains (DMS-V31, Gelest), cross-linker (HMS-301, Gelest), and catalyst (SIP6831.2, Gelest) are mixed thoroughly.  The mixture is degassed, and cured at $60^\circ\mathrm{C}$ for at least one week. By changing the ratio of chains to crosslinker, while keeping the concentration of catalyst consistent (at 0.0019 volume percent), the gel's Young modulus can be tuned.

To create `Sylgard' silicones, we  use the commercial brand Sylgard 184 (Dow).
The base is mixed with curing agent, using different ratios to attain different stiffnesses. The mixture is degassed thorougly and cured at $40-60^\circ\mathrm{C}$ for at least one week. For both silicones, the Young's modulus, $E$, is measured by indentation experiment \cite{styl15}.

\subsection{Measurement solute of $\phi_{sat}$ and D}
Silicone gels were prepared as a thin layer in 50mm diameter glass-bottom petri dishes (MatTek). We measured the mass of the gel with a microbalance before pouring a bath of fluorinated oil on top and allowing it to diffuse into the gel (either held at room temperature or at $40^\circ\mathrm{C}$).
Periodically, we removed the excess oil and recorded the mass of dissolved oil.
After around 30 hours for Gelest gels, and 60 hours for Sylgard gels, the weight plateaued as the gels reached saturation, allowing us to calculate $\phi_{sat}$ in vol\%.

\begin{figure}
 \centering
 \includegraphics[width=7cm]{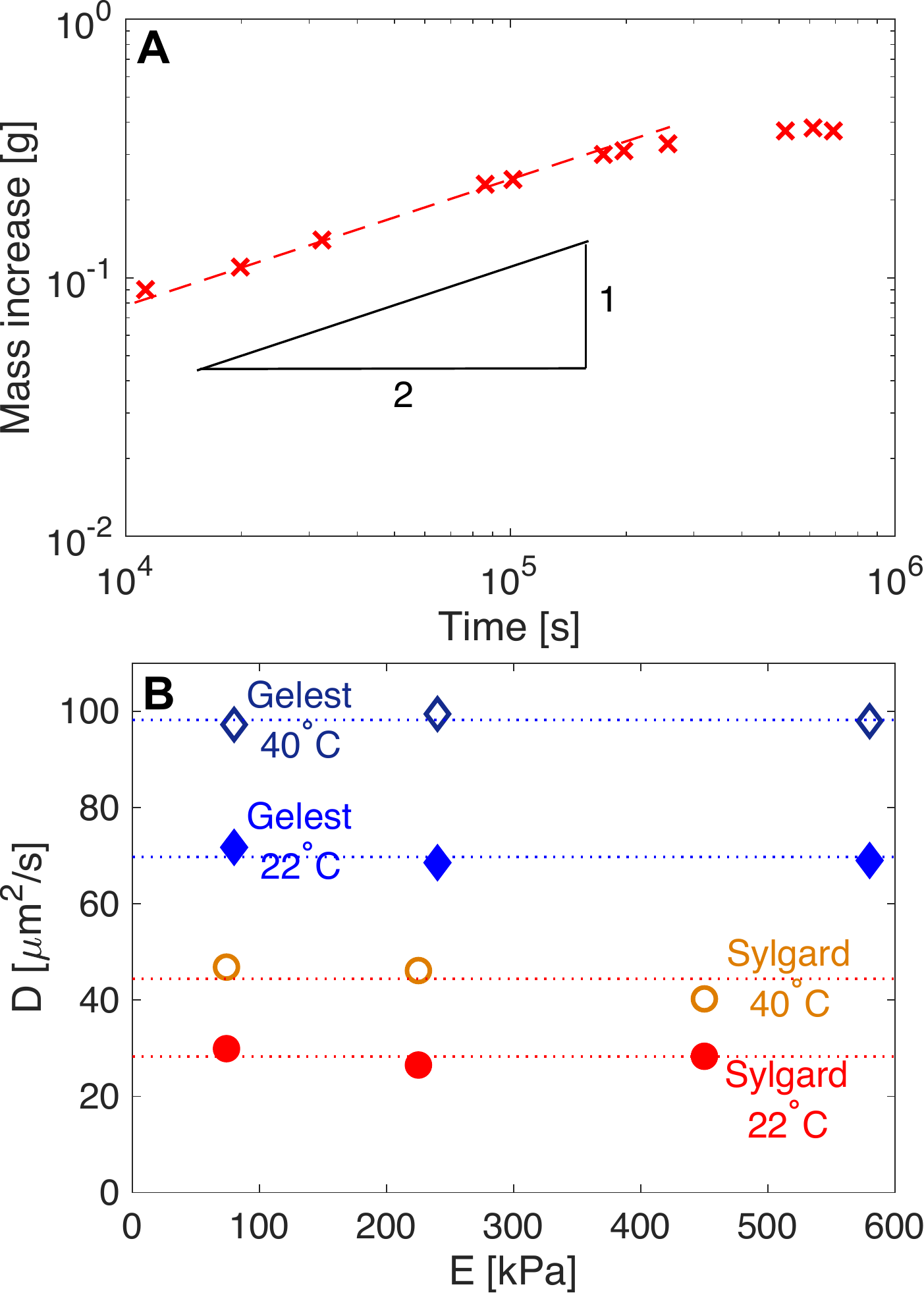} \\
 \caption{\emph{Measuring the diffusivity of fluorinated oil in silicone}. A) Here, Sylgard silicone with $E=230$kPa is immersed in fluorinated oil and weighed at regular intervals to determine the uptake in mass, as shown. Initially the uptake follows $m_{oil}\sim t^{1/2}$, as expected from equation (\ref{eqn:moil}). Subsequently, $m_{oil}$ plateaus at the saturation concentration. The red dashed line shows a fit with slope $1/2$. B) Plots of diffusivity vs $E$ show how $D$ is larger in Gelest silicones than in Sylgard silicones. $D$ is also independent of stiffness for the two silicone families, and increases with temperature.}
 \label{fig:saturation}
\end{figure}

The diffusion coefficient, $D$, was calculated following \cite{roso2020}.
Briefly, for an infinitely thick layer of silicone, covered in a layer of fluorinated oil, we expect the concentration profile (in terms of mass per volume) to follow
\begin{equation}
    c(z,t)=c_{sat}\mathrm{erfc}\left(\frac{z}{2(Dt)^{1/2}}\right),
\end{equation}
where $z$ is the distance from the interface.
This comes from solving the time-dependent diffusion equation, assuming saturation at the surface, so $c(z=0)=c_{sat}$, where $c_{sat}$ is the saturation concentration.
Integrating the concentration over the depth, $z$, gives the total mass of oil per unit area.
We then multiply by the area of the dish, to get the total mass of oil in the sample for each timepoint, $t$:
\begin{equation}
    m_{oil}(t) = 2 r_{\mathrm{dish}}^2 c_{eq}(\pi D t)^{1/2}
    \label{eqn:moil}
\end{equation}
where $r_{dish}$ is the radius of the petri dish.
Fitting the data from the initial stages of the experiments to Equation 2 (see Figure \ref{fig:saturation}), we find the diffusion coefficient, $D$, of fluorinated oil in each silicone gel.

\subsection{Numerical Simulations}
In order to simulate this system we adapted our elastic ripening model previously studied in \cite{vida2020} to account for materials with different saturation concentrations~$\phi_\mathrm{sat}$. 
Therefore, we describe the dynamics of droplets embedded in a diluted concentration field $\phi$. Each droplet is characterized by its position $\vec{x_i}$ and radius $R_i$ and follows the dynamical equation \cite{webe19} 
\begin{equation}
    \dfrac{\mathrm{d}R_i}{\mathrm{d}t}=\dfrac{D(\vec{x_i})}{\phi_\mathrm{in}R_i}\left[\phi(\vec{x_i})-\phi_\mathrm{cond}(\vec{x_i})\right],
    \label{eqn:droplet_dynamic}
\end{equation}
where $\phi_\mathrm{in}$ is the material concentration inside the droplets and $\phi_\mathrm{cond}$ is the equilibrium concentration of the dilute phase at a pressure $P$, which is given by \cite{roso2020}
\begin{equation}
    \phi_\mathrm{cond}=\phi_\mathrm{sat}\mathrm{exp}\left({P/n_L k_BT}\right).
\end{equation}
To derive Eq. \ref{eqn:droplet_dynamic} we assume that the gradients of $\phi_\mathrm{sat}$ are small in the vicinity of $\vec{x_i}$ and take only the leading order term. 
The conservation law for the solute in the dilute phase is given by \cite{vida2020}
\begin{equation}
    \partial_t\phi(\vec{x})=-\vec{\nabla} \cdot \vec{J}(\vec{x}) - \phi_\mathrm{in}\sum_i\dfrac{4\pi}{3}\dfrac{\mathrm{d}R_i^3}{\mathrm{d}t}\delta(\vec{x}-\vec{x_i}),  
    \label{eqn:dilute_phase_eqn}
\end{equation}
where the first term corresponds to fluxes driven by chemical potential differences according to Eq. \ref{eqn:J}, and the chemical potential $\mu$ is given by Eq. \ref{eqn:mu}. The second term on Eq. \ref{eqn:dilute_phase_eqn} asserts material conservation in exchange with the droplet phase. 
To model the transition region between the two materials, we use a sigmoidal function given by
\begin{equation*}
E=\dfrac{E_\mathrm{stiff}+E_\mathrm{soft}}{2}+\dfrac{E_\mathrm{stiff}-E_\mathrm{soft}}{2}\tanh{(x/\Delta)}
\end{equation*}
with $\Delta=5 \mu$m. The saturation concentration $\phi_\mathrm{sat}$ was model analogously following the same curve.
\\
We initialize the system by randomly placing droplets according to experimental density measurements \cite{styl2018}. The droplet radii are randomly drawn from a gaussian distribution following the data shown in Figure \ref{fig:sizes}. The dilute phase $\phi$ is initialized at $\phi=\phi_\mathrm{cond}$ to avoid initial droplet growth. In our simulations we assumed a densely packed droplet phase $\phi_\mathrm{in}=1$, other parameters are $\alpha=0.5$, $n_L k_BT=\unit[11]{MPa}$ and $\Upsilon= \unit[4]{mN/m^2}$. 

\subsection{Supplementary movies}

\textbf{Supplementary movie 1:} \textit{Elastic ripening from small to big}. Movie showing progression of the ripening experiment in Figure \ref{fig:ripening}a. Images are taken at 5 minute intervals.

\textbf{Supplementary movie 2:}  \textit{Elastic ripening from big to small}. Movie showing progression of the ripening experiment in Figure \ref{fig:ripening}b. Images are taken at 5 minute intervals.

\textbf{Supplementary movie 3:} \textit{Simulated elastic ripening from small to big.} 2-D Projection of a numerical simulation with Gelest ($\phi_\mathrm{sat}=0.023$ and $E=\unit[10]{kPa}$)  on the left, and Sylgard ($\phi_\mathrm{sat}=0.022$ and $E= \unit[80]{kPa}$) on the right. Chemical potential $\mu$ is shown as a density plot with colorbar at the right side.

\textbf{Supplementary movie 4:}  \textit{Simulated elastic ripening from big to small}. 2-D Projection of a numerical simulation with Gelest ($\phi_\mathrm{sat}=0.023$ and $E=\unit[140]{kPa}$)  on the left, and Sylgard ($\phi_\mathrm{sat}=0.022$ and $E= \unit[100]{kPa}$) on the right. Chemical potential $\mu$ is shown as a density plot with colorbar at the right side.

\section*{Conflicts of interest}
There are no conflicts to declare.

\section*{Acknowledgements}
The Acknowledgements come at the end of an article after Conflicts of interest and before the Notes and references.



%

\end{document}